\documentclass{article}

\usepackage{PRIMEarxiv}

\usepackage[utf8]{inputenc} 
\usepackage[T1]{fontenc}    
\usepackage{hyperref}       
\usepackage{url}            
\usepackage{booktabs}       
\usepackage{amsfonts}       
\usepackage{nicefrac}       
\usepackage{microtype}      
\usepackage{lipsum}
\usepackage{fancyhdr}       
\usepackage{graphicx}       
\graphicspath{{media/}}     

\usepackage{listings}
\usepackage{xspace}
\usepackage{multirow}
\usepackage{tabularx} 
\usepackage[nopostdot,nogroupskip,nonumberlist]{glossaries}

\title{Towards High Performance Quantum Computing (HPQ): Parallelisation of the Hamiltonian Auto Decomposition Optimisation Framework (HADOF)
}

\author{
  Namasi G. Sankar \\
  School of Computer Science\\
  Centre for Quantum Engineering,\\ Science, and Technology\\
  University College Dublin \\
  Dublin\\
  \texttt{namasivayam.gomathisankar@ucdconnect.ie} \\
   \And
  Georgios Miliotis\\
    Antimicrobial Resistance and\\ Microbial Ecology Group, \\ School of Medicine, \\ University of Galway,  \\Galway\\
  \texttt{georgios.miliotis@universityofgalway.ie} \\
  \And
  Simon Caton \\
  School of Computer Science\\
  Centre for Quantum Engineering,\\ Science, and Technology\\
  University College Dublin \\
  Dublin\\
  \texttt{simon.caton@ucd.ie} \\
}

\begin{document}
\maketitle

\begin{abstract}
Practical applicability of quantum optimisation on near term devices is constrained by limited qubit counts and hardware noise, which restricts the scalability of quantum optimisation algorithms for combinatorial problems. The simulation of large quantum circuits is also difficult and constrained by memory requirement. The Hamiltonian Auto Decomposition Optimisation Framework (HADOF) addresses this by decomposing large QUBOs into smaller subproblems that can be solved iteratively on quantum or classical backends. This allows the scalability of quantum QUBO algorithms beyond device limits, as well as their simulation on classical devices. In this research, we extend the evaluation of HADOF by benchmarking on real IBM QPUs across sequential, single-QPU parallel, and multi-QPU parallel execution modes, advancing toward High Performance Quantum (HPQ) computing for combinatorial optimisation problems.

Experimental results on IBM quantum hardware demonstrate up to 3-4× reduction in wall clock time when utilising four QPUs compared to sequential execution baseline, while maintaining comparable solution quality. Notably, even single QPU execution benefits from parallelised job orchestration and execution, yielding up to 3× speedup. Simulated results predict over 5x speed-up in parallel execution mode. We further validate the practical applicability of the approach on real world genome assembly instances, showing that both sequential and parallel HADOF variants achieve competitive accuracy while significantly improving time to solution. These results highlight the importance of parallelism at both the algorithmic and system levels, positioning HADOF as a viable pathway toward scalable quantum optimisation.
\end{abstract}


\newif\iffinal

\iffinal
  \newcommand\namasi[1]{}
  \newcommand\simon[1]{}
  \newcommand\georgios[1]{}
\else
  \newcommand\namasi[1]{{\color{red}[***Namasi: #1]}}
  \newcommand\simon[1]{{\color{blue}[***Simon: #1]}}
  \newcommand\georgios[1]{{\color{magenta}[***Georgios: #1]}}
\fi

\glsresetall
\section{Introduction}
\label{sec:intro}
 
Quantum computing has emerged as a promising paradigm for solving combinatorial optimisation problems that are intractable for classical algorithms, with applications spanning finance (risk analysis, prediction) \cite{qfin}, machine learning (hyperparamter tuning, feature selection) \cite{hypcomb, NAZIR2021102164}, and computational biology (genome assembly, sequence alignment) \cite{10.1093/bib/bbw096, contig_orientation}. Many such problems can be formulated as a Quadratic Unconstrained Binary Optimisation (QUBO), which maps naturally onto quantum Hamiltonians and can be addressed using near-term algorithms such as the Quantum Approximate Optimization Algorithm (QAOA) and Quantum Annealing (QA). However, the practical deployment of these methods remains severely constrained by the limitations of current quantum devices, particularly restricted qubit counts, circuit depth limitations, and hardware noise \cite{Gonzalez-Garcia2022-bn, Preskill2018-en}.

To address the scalability bottleneck, hybrid quantum-classical frameworks have been proposed that decompose large optimisation problems into smaller subproblems that can be executed on available hardware \cite{bravyi2020obstacles, qaoa2, kim2025distributed}. One such approach is the Hamiltonian Auto Decomposition Optimisation Framework (HADOF) \cite{Sankar2025HADOF} to solve Quadratic Unconstrained Binary Optimisation (QUBO) problems. QUBO is a generalised encoding strategy for many NP hard combinatorial optimisation problems which naturally maps into quantum Hamiltonians, making them suitable for encoding into both gate-based and annealing quantum computers. HADOF iteratively partitions a global QUBO Hamiltonian into sub-Hamiltonians that can be solved independently and recombined into a global solution. This enables problem sizes far exceeding the physical qubit capacity of current devices while maintaining competitive accuracy. Despite these advances, existing implementations of HADOF and similar decomposition methods are predominantly sequential, underutilising the parallelism that can be exploited after the decomposition. This limitation is increasingly significant as quantum computing moves toward distributed and hybrid architectures, where parallel execution across multiple quantum processing units (QPUs) and tight integration with High Performance Computing (HPC) resources are expected to play a central role \cite{HPQ}. 

Prior HADOF research was evaluated primarily through simulation and included a small proof-of-concept demonstration on IBM quantum hardware. In this paper, we introduce a parallelised extension of HADOF that enables concurrent execution of decomposed subproblems on a single QPU and across multiple QPUs, while also supporting large scale parallel simulation on classical CPUs. Our approach leverages Hamiltonian decomposition, pre-transpiled small width circuits, and asynchronous job orchestration to improve hardware utilisation and reduce scheduling overhead. Unlike traditional sequential workflows, the proposed framework exploits both inter-QPU parallelism and intra-QPU concurrency \cite{qiskit-ibm-runtime}, aligning with emerging paradigms in High Performance Quantum (HPQ) computing.
The proposed framework is evaluated on IBM quantum hardware, demonstrating substantial improvements in wall clock time (makespan) relative to sequential HADOF, achieving up to 3-4× speed-up when utilising multiple QPUs and 3x speed-up in single QPU concurrency. Furthermore, we validate the practical applicability of the approach on genome assembly instances \cite{quantum_genome_assembly}, highlighting its potential for scalable quantum optimisation in domain specific applications.
This work contributes to the growing body of research at the intersection of quantum algorithms, systems engineering, and HPC integration, and provides a concrete step toward scalable, parallel quantum optimisation in the Noisy Intermediate-Scale Quantum (NISQ) and early fault-tolerant eras.

\section{Related Work}
\label{sec:rw}

QAOA \cite{qaoa} and QA~\cite{Suzuki2009-fv} are foundational quantum methods for tackling QUBO problems. QAOA is a variational, gate-based algorithm alternating between cost and mixer Hamiltonians, generating a probability distribution over solutions favouring low-cost solutions~\cite{ising-qubo, kim2025distributed}. QA, on the other hand, is an analogue adiabatic method evolving a quantum system from an initial to the problem Hamiltonian. Both produce biased sample distributions over candidate solutions, offering practitioners flexibility when selecting among high-quality alternatives. However, NISQ hardware limits QAOA/QA to small problem sizes due to qubit and connectivity constraints~\cite{kim2025distributed}. This motivates hybrid and decomposition approaches.

In classical optimization, divide-and-conquer and decomposition heuristics are standard for scaling to large problems. General purpose solvers (e.g., IBM CPLEX~\cite{docplex2024}) can struggle QUBO problems beyond hundreds of variables~\cite{NP-hard-qubo}, motivating decomposition and hybrid quantum-classical methods. The multilevel QAOA of Maciejewski \textit{et al.}~\cite{maciejewski2023multilevel} splits a large QUBO into manageable sub-QUBOs that are solved iteratively or in parallel and then recombined. These techniques enable practical scaling and lay the foundation for distributed quantum optimization. Recent strategies distribute or decompose QAOA across subproblems. Recursive QAOA (RQAOA)~\cite{bravyi2020obstacles} uses QAOA to iteratively fix qubits and shrink the problem, focusing quantum resources on the hardest sub-instances. The QAOA-in-QAOA (QAOA$^2$) and related parallel QAOA heuristics~\cite{qaoa2} decompose a large graph (e.g., MaxCut) into subgraphs, solve each with QAOA in parallel, and merge the results, exploiting HPC for scalability. Early approaches worked best on sparse or weakly coupled problems~\cite{kim2025distributed}, but dense QUBOs require advanced coordination to manage strong variable interactions. The Distributed QAOA (DQAOA) framework~\cite{kim2025distributed} extends parallelization further. Large QUBOs are decomposed into sub-QUBOs, solved on quantum or classical resources in parallel, with an aggregation policy reconciling overlaps and correlated interactions. This iterative approach scales to large, dense QUBOs; for example, Kim et al. report around 99\% approximation ratios on 1,000-variable instances within minutes, outperforming prior methods in both quality and time-to-solution. DQAOA leverages quantum-centric HPC platforms to update a global solution iteratively, demonstrating that distributed computing augments quantum optimization for practical problem sizes. HADOF and DQAOA overcome standard QAOA scalability limits via decomposition. DQAOA relies on explicit partitioning and parallel aggregation, excelling on HPC or distributed platforms. HADOF uses adaptive, iterative refinement with a probabilistic global view, reducing quantum requirements per step. While DQAOA is I would say designed to exploit federated parallelism for raw parallelism and wall-clock minimization, HADOF provides efficient sequential scaling and solution diversity. Both frameworks represent the cutting edge of distributed quantum optimization, and hybrid approaches combining their strengths are promising future directions.

Previously, HADOF was tested for scalability using classical simulations on Pennylane \cite{pennylane} circuits without noise. In this research, we test HADOF on real IBM QPUs to measure wall clock time and accuracy on currently available hardware. Additionally, we compare HADOF on its accuracy and speed when executed in a sequential and parallelised manner. We also show the practical usability of QUBO and HADOF based optimisation for genome assembly on a sample dataset.

\section{Application Scenario} \label{sec:applications}

\subsection*{Genome Assembly as a Travelling Salesman Problem}

The bioinformatics domain inherently includes many NP-hard combinatorial optimization problems. In this paper, we focus on genome assembly recast as a combinatorial problem. Genome assembly is a fundamental task in genomics, aiming to reconstruct an organism's complete DNA sequence from sequencing reads. Reads are the short fragments of DNA sequence data that come directly out of a sequencing machine. They capture the order of nucleotide base pairs (bp) A, C, G, and T from a piece of a genome but do not represent the whole genome in one long contiguous stretch \cite{Wajid2012-sy}. Sequencing read fragments need to be assembled in the correct order to reconstruct the original DNA sequence. Accurate and timely genome assembly has significant societal impact. In infectious disease surveillance, whole genome sequencing enables rapid identification of pathogens, tracing of transmission pathways, and detection of emerging variants, which are critical for effective outbreak response \cite{Gilchrist2015-hs}. Similarly, in cancer genomics, sequencing is essential for identifying mutations, structural variations, and virus–host integrations that influence disease progression and treatment strategies \cite{Bergmann2025-xy}. Modern genome assembly algorithms typically represent reads using overlap graph based models in which nodes correspond to reads or sequence fragments and edges represent overlaps between sequences. Overlaps between reads are first identified, a graph describing read relationships is constructed, and a consensus sequence is generated from a traversal of the resulting graph. In overlap graphs, reads are represented as vertices and edges correspond to suffix–prefix overlaps. Recovering the genome therefore requires identifying a path that visits reads in the correct order. Under common formulations, this corresponds to solving a TSP on the overlap graph \cite{shomorony2016information}. 

Since this problem is NP-hard, no polynomial time algorithm is known for solving it in the general case. Because of this complexity, most practical assemblers rely on heuristics and graph simplification techniques \cite{Wick2017}. The combinatorial nature of genome assembly therefore motivates exploring alternative optimisation paradigms capable of efficiently searching large solution spaces. Quantum computing has emerged as a potential framework for tackling such optimisation problems. Recent work has explored quantum computing as a new paradigm for solving such optimization problems \cite{boev_qubo_assembly_2021} \cite{sarkar_quaser_2021}. QAOA \cite{qaoa} and Quantum Annealing \cite{Kadowaki1998} aim to use superposition and tunnelling to explore complex solution spaces more efficiently than classical heuristics, potentially reducing time-to-solution for hard combinatorial subproblems in assembly in the future \cite{Pamidimukkala:2026sny}. The assembly problem can be reformulated as optimising paths through a graph with constraints (read ordering, overlap consistency), which is conceptually similar to the Travelling Salesman Problem (TSP). To encode the problem onto quantum Hamiltonians, they are transformed into QUBO with general form expressed as:

\begin{equation}
\begin{split}
\min_{x} \mathrm{x}^T Q \mathrm{x} 
= \min_{x} \biggl( \sum_{i} \sum_{j \ge i} Q_{ij} \, x_i \, x_j \biggr) \\ 
= \min_{x} \biggl( \sum_{i} Q_{ii} \, x_i 
  + \sum_{i} \sum_{j > i} Q_{ij} \, x_i \, x_j \biggr)
\end{split}
\end{equation}

where $Q \in \mathbb{R} ^{n\times n}$ is an upper triangular matrix and $\mathrm{x}$ is a vector of binary variables \cite{Boros2007-zp, Wang2009-lt}. In genome assembly, overlaps between sequencing reads are computed, forming a directed graph. Each read is treated as a node, and overlaps define edges. Let a directed graph be given in the form $G=(V,E)$, where $V=\{1,2,\ldots,N\}$ is a set of vertices, and $E$ is a set of directed edges consisting of pairs $(u,v)$ with $u,v\in V$. The solution of the Hamiltonian path problem can be represented by an $N\times N$ permutation matrix $\mathcal{X}=(x_{v,i})$, whose unit elements $x_{v,i}$ indicate that vertex $v$ is visited at step $i$. We assign each element $x_{v,i}$ of the matrix $\mathcal{X}$ a separate binary logical variable within the QUBO optimization problem. Note that this representation results in a polynomial overhead in the number of logical variables: solving the Hamiltonian path problem for an $N$-vertex graph requires $N^2$ logical variables. The resulting Hamiltonian for the corresponding QUBO problem takes the form

{\fontsize{8}{11}\selectfont
\begin{equation}
\begin{split}
\mathcal{H} = A \sum_{v=1}^N \left(1 - \sum_{j=1}^N x_{v,j}\right)^2
+ A \sum_{j=1}^N \left(1 - \sum_{v=1}^N x_{v,j}\right)^2 \\ + A \sum_{(u,v)\notin E} \sum_{j=1}^{N-1} x_{u,j} x_{v,j+1}    
\end{split}
\end{equation}
}

where $A>0$ is a penalty coefficient. The first two terms enforce that each vertex appears exactly once in the path and that each position in the path is occupied by a single vertex. The third term penalizes transitions between vertices that are not connected by an edge in the overlap graph. In general, the Hamiltonian path mapping described above applies to both cyclic and acyclic directed graphs. However, in ideal cases the overlap graph contains no cycles \cite{boev_qubo_assembly_2021}. The target objective is path recovery rather than a closed tour. In such cases, the mapping can be simplified to reduce the qubit overhead. For a directed acyclic graph (DAG) $G=(V,E)$, define a set of binary variables $\{x_{u,v}\}_{(u,v)\in E}$, where $x_{u,v}=1$ indicates that the edge $(u,v)$ is included in the Hamiltonian path. In this representation, only $M=|E|$ logical variables are required, with $M<N^2$ in typical acyclic graphs. The corresponding Hamiltonian can be written as

{\fontsize{8}{11}\selectfont
\begin{equation}
\mathcal{H} = A \sum_{u\in V} \left(1 - \sum_{(u,v)\in E} x_{u,v}\right)^2 
+ A \sum_{v\in V} \left(1 - \sum_{(u,v)\in E} x_{u,v}\right)^2
\end{equation}
}
\vspace{0.25cm}

where $A>0$ is a penalty coefficient. The first term enforces that each vertex has one outgoing edge in the path, and the second term enforces one incoming edge. This compact encoding requires only $M$ logical variables corresponding to edges in the acyclic graph and thus reduces the overhead compared to the full permutation matrix representation.

\section{Methods}

\subsection*{\textbf{General Overview}}

HADOF proceeds iteratively as follows:
\begin{enumerate}
    \item Encode the full QUBO as a Hamiltonian.
    \item Apply an optimisation algorithm (like QAOA or QA) that produces a probability distribution to sample approximate solutions.
    \item Approximate sub-Hamiltonians using the expected values of the binary variables (qubits).
    \item Solve each sub-Hamiltonian iteratively.
    \item Aggregate sampled solutions from sub-Hamiltonians to guide the next iteration.
\end{enumerate}

In the parallel approach, instead of solving each sub-Hamiltonian iteratively in step 4, we generate all the sub-Hamiltonians at the beginning of each HADOF iteration in an independent manner, and then solve them in parallel. This way, each iteration is made of independent sub-circuits that can be solved asynchronously and then aggregated together. 

\subsection*{\textbf{Implementation Details}}

Let the global problem be represented in the standard QUBO form:
\begin{equation}
\min_{x \in \{0, 1\}^n} \quad x^T Q x
\end{equation}
where $Q \in \mathbb{R}^{n\times n}$ is an upper triangular cost matrix, and $n$ is the dimensionality of the binary decision variable $x$. The corresponding quantum Hamiltonian $H_Q$ encodes the QUBO in the computational basis. To scale the optimisation process, $H_Q$ is decomposed into a set of sub-Hamiltonians, each defined over a subset of the full variable set. Let $S_i \subset \{x_1,...,x_n\}$ denote the variables of subproblem $i$, with $\|S_i\| = k<<n$. The sub-Hamiltonian $H_i$ is defined by: 

\begin{equation}
H_i = \mathbb{E}_{x_{\bar{S}_i}} = P ( H_Q \big| x_{S_i}, x_{\bar{S}_i})
\end{equation}

\begin{figure}
    \centering
    \includegraphics[width=1\linewidth]{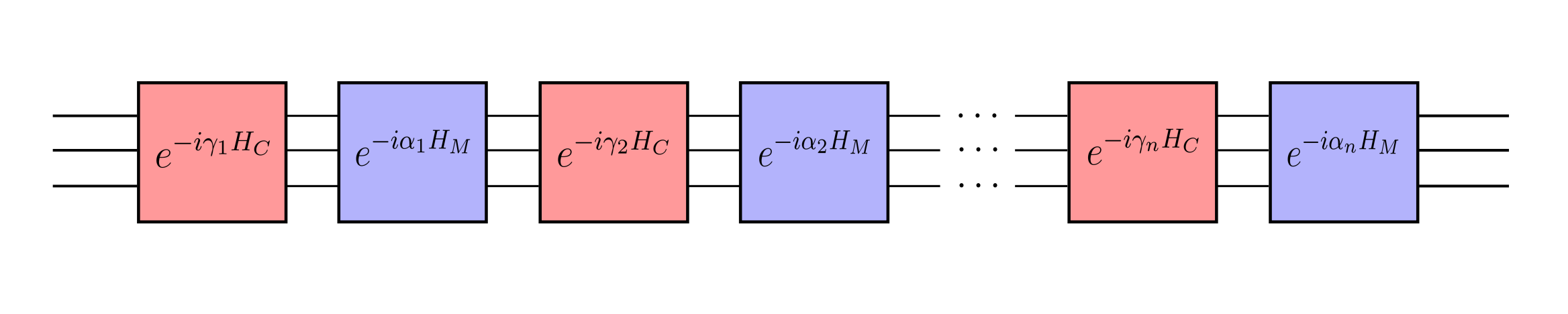}
    \caption{Standard QAOA circuit with alternating cost and mixer Hamiltonians\cite{QUBO-pennylane}. The output produces a probability distribution over the solution space which can be samples with the shots parameter of the quantum simulation or backend. Higher sampled solutions are more likely to be solutions with better objective value.}
    \label{fig:QAOAcircuit}
\end{figure}

Here, $\bar{S}_i$ denotes the complement of $S_i$, and $P(x_{\bar{S}_i})$ is a distribution over unsampled variables.  We use \( \mathbb{E}[x_k] = P(x_k) \) as the expected value that variable \( x_k \) is 1, estimated from previous iterations or prior knowledge. Ideally, this expectation is estimated using a weighted average over all states the rest of the QUBO can assume. In this study, \( \mathbb{E}[x_k] \) is approximated as the expected value of each qubit, by sampling it. This transformation embeds global context into each subproblem while keeping the computational cost tractable. To estimate $\mathbb{E}[x_i]$, we use digitesed QAOA \cite{trotter_digitized_annealing_qaoa}. For each subproblem \( i \), a QAOA circuit is constructed using the cost and mixing unitaries, as in Fig. \ref{fig:QAOAcircuit}:

\begin{align}
U(H_i, \gamma) &= e^{-m \gamma_m H_i} \\
U(M, \beta) &= e^{-m \beta_m \sum_{j=1}^{k} X_j}
\end{align}

\begin{figure}
    \centering
    \includegraphics[width=.9\linewidth]{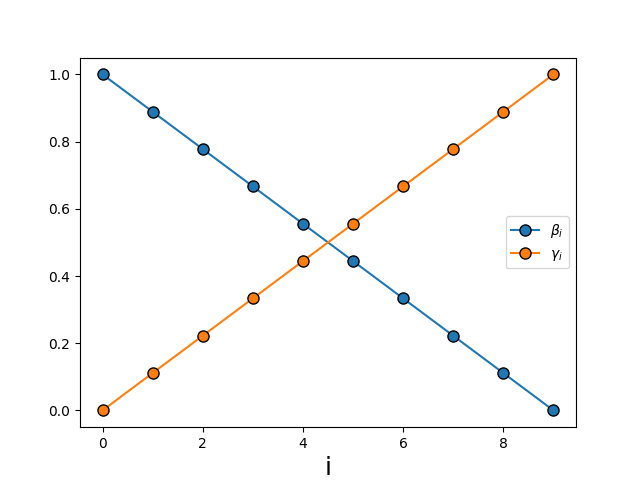}
    \caption{Trotterised QAOA parameters based on Ref \cite{QUBO-pennylane}, \cite{openQAOA}.  We move $\beta_m$ from 1 to 0 and $\gamma_m$ from 0 to 1 allowing the system to stay close to the ground state of the mixer Hamiltonian to the Cost Hamiltonian.}
    \label{fig:qaoaparams}
\end{figure}

Here, we implement QAOA as a trotterisation of QA, using the Annealing Parametrisation \cite{openQAOA}, to avoid the classical optimisation loop required to find $\beta_m$ and $\gamma_m$. We start in the ground state $\left| + \right\rangle^{\otimes k}$ of the mixer Hamiltonian $X$ and move to the ground state of the cost Hamiltonian $H_i$ slowly enough to always be close to the ground state of the Hamiltonian, as in Fig. \ref{fig:qaoaparams}. This rate is determined by the number of layers $p$. We initialize the $\beta_m$ and $\gamma_m$ in this way, moving $\beta_m$ from 1 to 0 and $\gamma_m$ from 0 to 1. 

\begin{figure*}
    \centering
    \includegraphics[width=.95\linewidth]{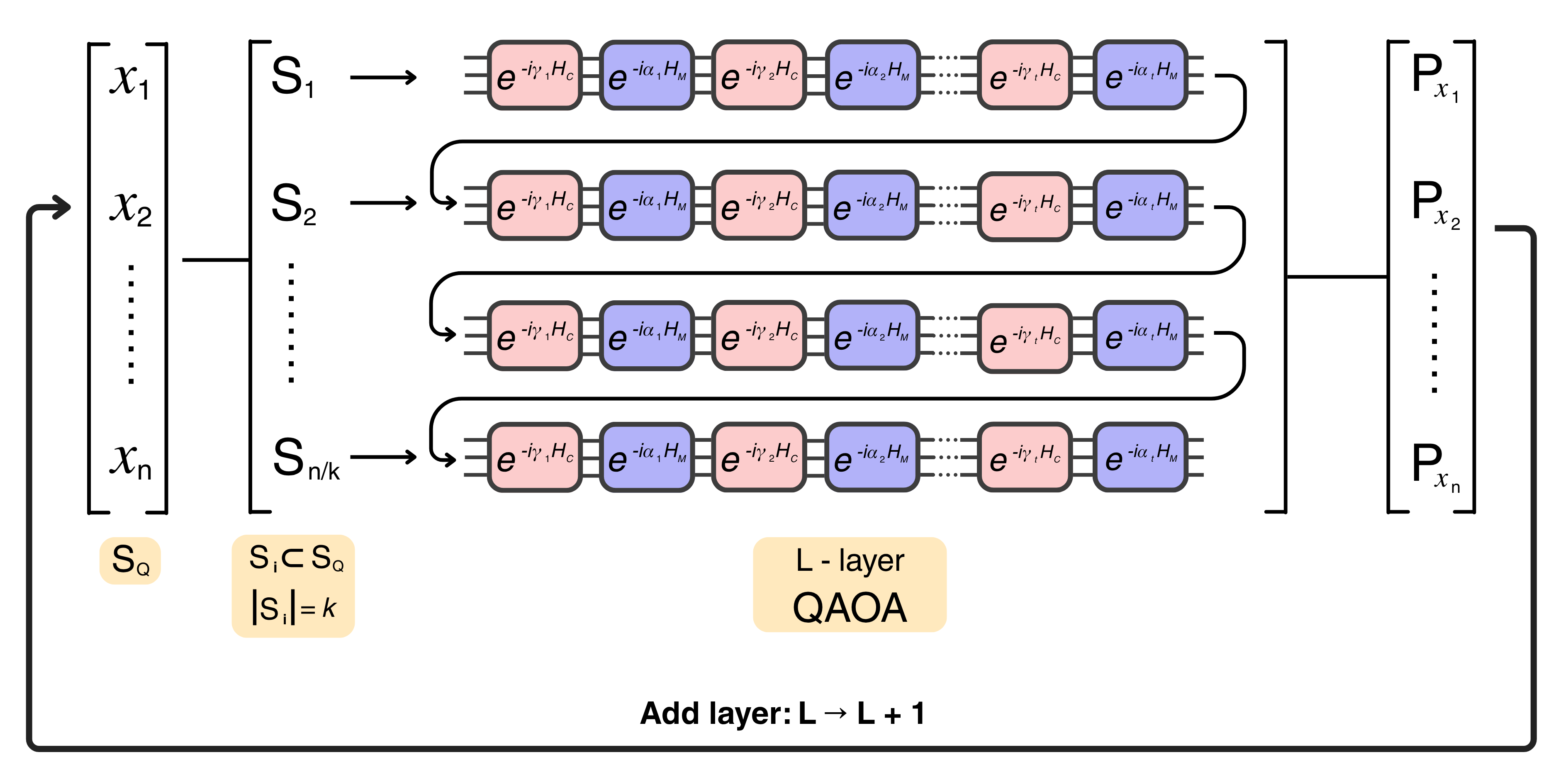}
    \caption{General overview of the HADOF framework. Here, we use QAOA as the optimiser, which is called iteratively. 
    \textbf{(1)} We choose subsets of sizes 5 from the binary variables of the global problem. 
    \textbf{(2)} These are used to form the sub-Hamiltonians using $P(x_i)$, approximated as the expected value of each qubit. 
    \textbf{(3)} The QAOA circuit set up with $t=1$ layers and in every iteration we add a layer. In this study, we use 5 layers. 
    \textbf{(4)} Once the $n/k$ sub-Hamiltonians are optimised, we sample them and use an aggregation policy to form the global solution probability distribution. This can be done sequentially or in parallel.}
    \label{fig:HADOF}
\end{figure*}

However, instead of applying the QAOA procedure completely, two changes are made to iteratively estimate the sub-Hamiltonians. 1) In each iteration of the loop, every sub-Hamiltonian is solved using QAOA, however, the whole circuit is not applied. In the $l^{th}$ iteration, only layers 1 to $l$ are applied. 2) To approximate the value of $\mathbb{E}[x_i]$, individual qubits are sampled instead of sampling from all possible solutions of the QAOA in every iteration. The average for each qubit is used as a proxy for $\mathbb{E}[x_i]$. The optimisation process unfolds over \( p \) global iterations as in Fig. \ref{fig:HADOF}.

As seen in Fig. \ref{fig:HADOF}, in every iteration, a layer is added. In the sequential implementation of HADOF, within each layer (iteration), each sub-Hamiltonian is solved, then the expected value, $\mathbb{E}[x_i]$, is updated, before the next sub-Hamiltonian is generated. In the parallel version, we generate all the sub-Hamiltonians at the beginning of each iteration. Instead of updating the new expected values, the sub-Hamiltonians are generated using the values from the previous iteration. Note that at the beginning all the expected values are set to 0.5. Once all the sub-Hamiltonians of the iteration are solved, all of the values are updated before the next iteration begins.

\subsection*{\textbf{Parametric Details}}

\begin{enumerate}
    \item \textbf{Initialisation:}
    Create a global probability vector \( P(x_k) \in \mathbb{R}^n \), with all values initialised to 0.5, except for the first subset $S_i$
    
    \item \textbf{QAOA Loop:} 
    For \( L = 1, 2, \dots, p \):
    \begin{enumerate}
        \item Initialise QAOA circuit with $l$ layers with the first $l$ values of $\beta_m$ and $\gamma_m$
        \item For each model \( i \in \{0, \dots, M-1\} \):
        \begin{itemize}
            \item Replace inactive variables $\bar{S}_i$ by fixed expected values from previous iterations
            \item Construct sub-QUBO for subset \( S_i \) using the expected values $P(x_{\bar{S}_i})$ from previous iteration.
            \item Convert to Ising form: \( Q \to (h, J) \) and get the sub-Hamiltonian corresponding to the sub-QUBO
            \item Apply QAOA circuit of depth \( L \) on \( k \) qubits where \( k = \|S_i\|\) :
            \item Update \( P(x_k) \) vector by measure expected values of each qubit for current model:
            \begin{equation}
            P(x_i) = \mathbb{E}[x_i] 
            \end{equation}
        \end{itemize}
    \end{enumerate}
    
    \item \textbf{Final Output:} 
    After the final iteration, run the QAOA circuits with full depth (all layers in the beta schedule) to extract binary samples. Collect and store final solutions from all models. These solutions and their probabilities can be aggregated to form the global solution.
\end{enumerate}

The QAOA loop in step 2 can be run sequentially or in parallel. When run sequentially, the construction of sub-QUBO for subset \( S_i \) takes into account the updated expected values $P(x_{\bar{S}_i})$ of previously solved models within the current iteration. For the models that have not been solved yet, it uses the expected values from the previous iteration. For the parallel version, the expected values for every model (sub-Hamiltonian) is from the previous iteration, so that the problems can be generated independently. To parallelise the loop, we use joblib \cite{joblib}. $P(x_i) = 0.5$ is used as initialisation for all $i$. Circuits use 5 layers with $\beta_m = 1-(m/5)$ and $\gamma_m = m/5$. After each sweep of the $n/k$ circuits, we add one layer. To measure the individual qubits to update $P(x_i)$ we use 500 shots per qubit measurement. Finally, each circuit is sampled over all $k$ qubits using 5000 shots per circuit, to produce a distribution over each sub-solution. 5,000 global solutions are formed by concatenating sampled sub-solutions in sampling order and then evaluate their objective values. 

For banchmarking, random QUBO problems were generated by filling an upper triangular matrix using a uniform random number generator between -10 and 10. We present comparisons of accuracy and speed with Simulated Annealing from D-Wave samplers \cite{dwave-ocean-sdk} for problems with $n = 100, 200, ..., 500$ binary variables. We choose \( k = 5\) as the sub-Hamiltonian size because the accuracy of circuits degrades as the circuit size increases. The number of QAOA circuits per iteration will be \( n/k \). We also present the comparison in accuracy and speed of classically simulated noiseless models, classically simulated noisy model, and HADOF on real QPU in both sequential and parallel modes. However, due to limited time availablity and the cost of running real device experiments, we show feasibility and benchmarking on the real device only for 100, 200 and 300 binary variables. All the circuits are designed using Qiskit \cite{qiskit-ibm-runtime}. All the classical simulations were run on Macbook Pro M3 using Qiskit. Parallelised simulations were run with 10 cores on single threads. Ideal quantum simulations were performed on the Aer Simulater and noisy simulations were performed using the Fake Torino model. IBM\_Kingston was used for sequential and parallelisation on a single QPU experiments. On this Qiskit platform, the scheduler automatically transpiles runs queued jobs in parallel on a single device where possible \cite{qiskit-ibm-runtime}. Parallelisation across multiple QPUs made use of ibm\_kingston (Heron r2 processor), ibm\_pittsburgh (Heron r3), ibm\_fez (Heron r2) and ibm\_marrakesh (Heron r2). Multi-QPU execution is achieved by submitting jobs separately to the different QPUs. Fake Torino is a 133 qubit simulater. IBM Kingston, Pittsburgh, Fez and Marrakesh are 156 qubit devices. All devices posses a heavy-hex lattice topology, and were used in April, 2026.

\section{Results and Discussion}
\label{sec:results}

For each problem size, we provide a randomly generated QUBO problem to benchmark accuracy and speed. Each experiment was repeated 5 times and averaged for every problem size, except for real device executions, which were only performed once, due to the cost of computation and finite QPU budget. This is followed by experimentation using a genome assembly dataset (as discussed in Section \ref{sec:applications}).  

\subsection{Speed}

\begin{figure}
    \centering
    \includegraphics[width=.9\linewidth]{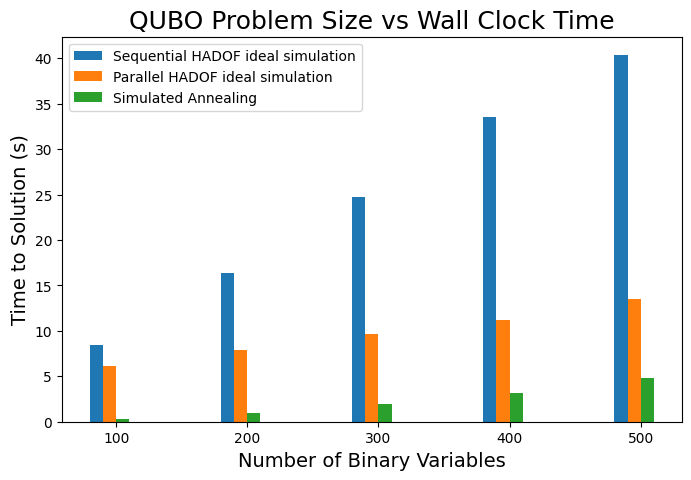}
    \caption{Simulating quantum circuit dynamics is hard classically, however with HADOF, problems up to size 500 variables are simulated in a comparable time to classical Simulated Annealing. Time advantage can be observed in HADOF simulation when running in parallel.}
    \label{fig:ttsclassical}
\end{figure}

We measure the time-to-solution using two metrics - wall clock time (makespan) and QPU usage time. Wall clock time is the duration from the beginning of execution to the end, while the QPU usage time measures the cost of computation on real device. On real device, the queuing time for submitted jobs is inevitable. QPU usage time sums up the time for every individual circuit job submitted, without the queuing time. To maintain fair benchmarking, we chose to run the simulations after noting down the pending jobs as shown on the Qiskit platform, and to our best knowledge, we began submitting jobs when the pending jobs was 0. The QPU usage is easy to predict on real device. Each 5 qubit circuit takes 3 seconds to run. For a problem size $n$, we make use of $n/5$ circuits per iteration (layer). We limit our circuits to 5 layers, which gives us 5 iterations. QPU usage for executing HADOF for an $n$ variable problem takes $3*n$ seconds. This does not change when run in parallel; QPU usage is calculated per job. 

\begin{figure}
    \centering
    \includegraphics[width=.9\linewidth]{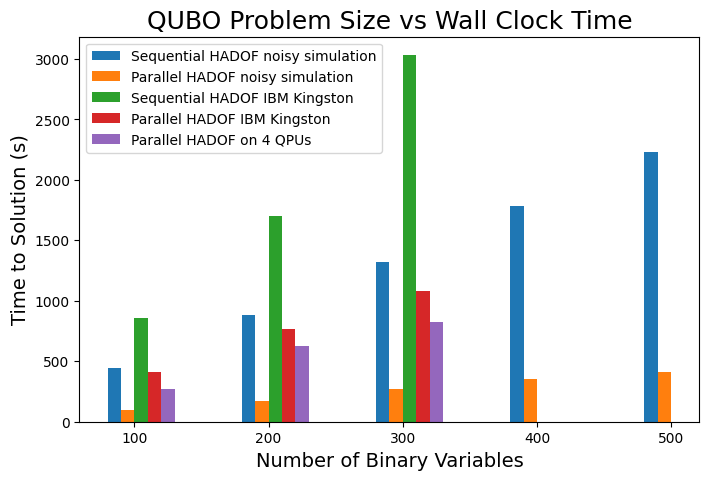}
    \caption{Note up to 5 times faster simulation speed when run with noise. On a real device, additional to run time, queueing time is significant. Submitting and running jobs in parallel saves both QPU usage and wall clock time.}
    \label{fig:ttsquantum}
\end{figure}

Fig. \ref{fig:ttsclassical} and Fig. \ref{fig:ttsquantum} show the wall clock time to solution for the different algorithms. In the ideal simulation setting (Fig.~\ref{fig:ttsclassical}), both sequential and parallel HADOF exhibit approximately linear scaling with respect to the number of binary variables. However, the parallel implementation consistently achieves lower wall clock times, with the performance gap widening as the problem size increases. For instance, at 500 variables, parallel HADOF achieves roughly a $3\times$ speedup over the sequential variant. This indicates that parallelisation not only reduces constant overheads but also improves the effective scaling coefficient, suggesting strong suitability for large-scale QUBO instances. In contrast, simulated annealing demonstrates significantly lower wall clock times across all problem sizes, reflecting its efficiency as a classical heuristic. Simulating (ideal and noisy simulations) are more computationally expensive in comparison. On the IBM QPUs currently, speed is still limited. The goal is that future devices will be able to solve larger problems with competent accuracy enabling larger decomposition sizes instead of 5 qubit circuits as shown currently, which might reduce wall clock time and QPU usage drastically. In Table \ref{tab:table}, note that running 10-50 qubit problems in a single iteration still only takes 3 seconds, which shows that QPU usage is not majorly affected by the size of the circuit. This suggests that if future QPUs feature larger and higher fidelity subcircuits, HADOF decomposition may provide a quantum advantage become increasingly competitive for large QUBO instances. 

Under noisy simulation (Fig.~\ref{fig:ttsquantum}), the wall clock time increases substantially for all HADOF variants, with sequential execution experiencing the most pronounced degradation. Notably, parallel HADOF mitigates this overhead effectively, achieving consistent speedups of approximately 4-5x over its sequential counterpart across problem sizes. This highlights that parallelisation becomes increasingly critical in realistic quantum computing regimes, where noise and execution overhead dominate QPU usage. Comparisons with IBM hardware (Kingston backend) further reveal that real device execution incurs significantly higher wall clock times than noisy simulation, likely due to queue delays, limited circuit throughput, and hardware constraints. Parallel execution on the hardware significantly improves performance relative to sequential execution by around 3x. Additionally, the multi-QPU configuration demonstrates slightly improved performance compared to parallelisation on a single QPU, indicating partial but non-ideal scaling, likely due to communication and orchestration overheads by the QPU scheduler. These results demonstrate that parallel HADOF provides substantial QPU usage advantages, particularly in noisy and current quantum hardware environments. The observed scaling behaviour supports the use of parallel and distributed quantum optimisation frameworks as a pathway toward high-performance quantum computing (HPQ), where mitigating execution overhead is essential for practical scalability.

\subsection{Accuracy}

\begin{figure}
    \centering
    \includegraphics[width=.9\linewidth]{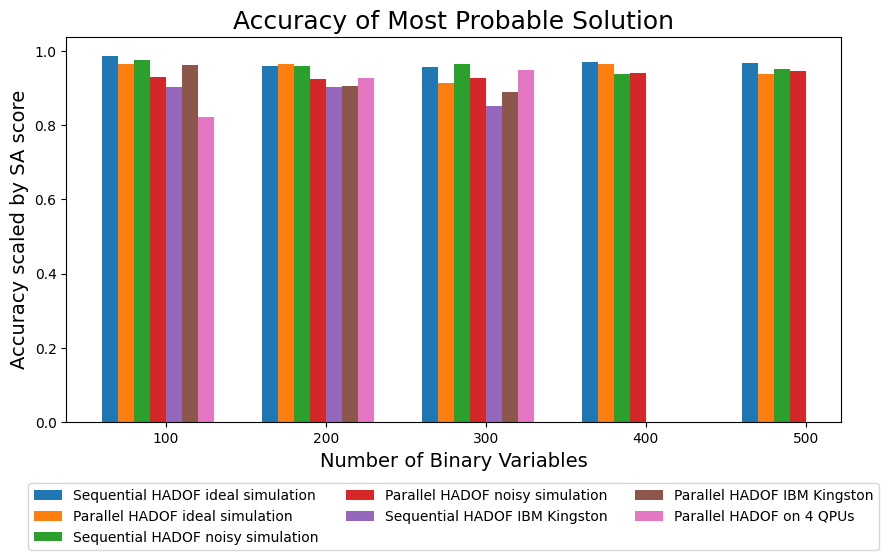}
    \caption{Most probable solution from all the algorithms is always above 82\% compared to classical Simulated Annealing.}
    \label{fig:accb}
\end{figure}
 
The accuracy of the obtained solutions is presented under two complementary metrics: (i) the accuracy of the most probable sampled solution, and (ii) the average accuracy across the sampled solution distribution. In both cases, solution quality is evaluated using the QUBO objective function and normalised with respect to the solution obtained via classical simulated annealing, which is assigned a score of 1. This normalisation enables consistent comparison across methods, particularly at larger problem sizes where exact solvers become impractical. From Fig.~\ref{fig:accb}, it is observed that under ideal simulation, both sequential and parallel HADOF achieve near-optimal performance, with accuracy consistently above $0.95$. The gap between sequential and parallel implementations is minimal in this regime, indicating that parallelisation does not degrade solution quality in noiseless conditions.

In the presence of noise (fake backend simulation), a slight degradation in accuracy is observed, particularly for the parallel variant at smaller problem sizes. However, both sequential and parallel noisy implementations remain within a narrow band ($\sim 0.92$--$0.97$), demonstrating robustness of the optimisation process to moderate noise levels. On real hardware, the impact of noise and device constraints becomes more pronounced, but remains competetive consistently maintained above 0.80. Sequential execution on the IBM backend exhibits a noticeable drop in accuracy, in some cases falling below $0.90$. Parallelisation across multiple QPUs demonstrates competitive relative to single device execution. 

\begin{figure}
    \centering
    \includegraphics[width=.9\linewidth]{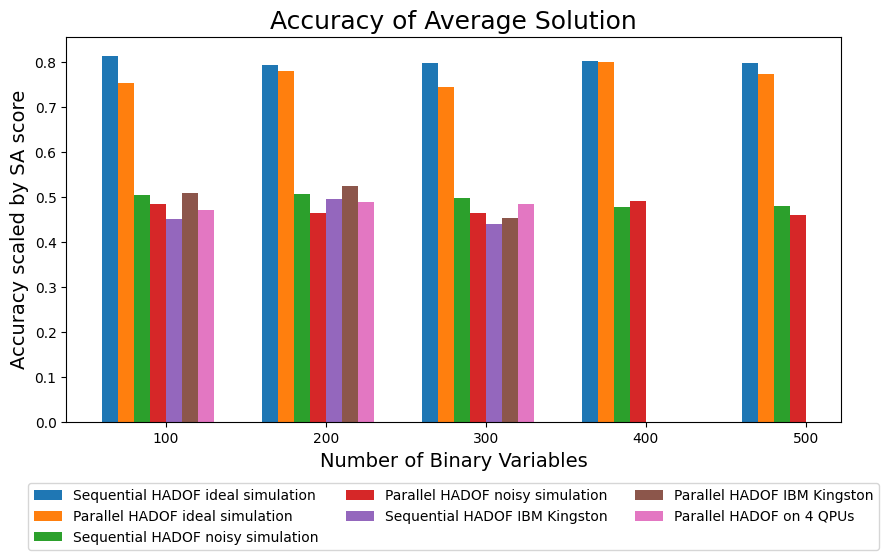}
    \caption{Average accuracy of the solutions from the distribution produced is between 70\% and 80\% of Simulated Annealing. However, on a real device this accuracy lowers to around 50\%.}
    \label{fig:acca}
\end{figure}

Fig.~\ref{fig:acca} reveals a more distinct separation between methods when considering the average accuracy across all sampled solutions. While ideal simulations still maintain high performance, the average accuracy is consistently lower than the most probable solution accuracy, indicating that suboptimal solutions contribute non-negligibly to the sampled distribution. Under noisy simulation, the degradation in average accuracy is more significant than in the most probable case, highlighting the sensitivity of the overall solution distribution to noise. Sequential implementations tend to retain slightly higher average accuracy than their parallel counterparts, suggesting that while parallelisation improves QPU usage, it may introduce greater solution variance. On real hardware, average accuracy decreases substantially compared to ideal and noisy simulations, in some cases dropping below $0.60$ for larger problem sizes. This indicates that noise and hardware limitations significantly affect the quality of the full solution distribution, even when the best sampled solution remains relatively strong.

Taken together, these results highlight a key distinction between the two metrics. The most probable solution accuracy remains relatively robust across all settings, suggesting that HADOF is capable of reliably identifying high-quality solutions even in noisy and hardware-constrained environments. In contrast, average accuracy reveals a broader degradation in the solution landscape, indicating that noise primarily affects the distribution of sampled solutions rather than the peak performance. Importantly, parallelisation does not significantly compromise the quality of the solutions found in any of these cases, while providing substantial QPU usage benefits (as shown in previous sections). Overall, the results demonstrate that HADOF maintains strong solution quality relative to classical baselines, with parallel implementations offering a favourable balance between computational efficiency and accuracy, particularly in realistic quantum computing regimes.

\subsection{HADOF against standard QAOA}

\begin{table*}[t]
    \vspace{-1em}
  \centering
  \caption{Comparing HADOF vs Standard QAOA}
  \label{tab:table}
  \begin{tabular}{|c|l|c|c|c|c|}
    \hline
    \textbf{Problem Size} & Algorithm & Wall Clock Time & QPU Usage Time & Best Accuracy & Average Accuracy \\
    \hline

    \multirow{4}{*}{10} 
    & Ideal Full     & 0.6  & -  & 0.94 & 0.12 \\
    & Ideal HADOF    & 1.31 & - & 1 & 0.76 \\
    & Kingston Full  & 16.66 & 3   & 1 & 0.21 \\
    & Kingston HADOF & 85.3 & 30   & 1 & 0.48 \\
    \hline

    \multirow{4}{*}{20} 
    & Ideal Full     & 1.51  &  - & 0.83 & 0.08 \\
    & Ideal HADOF    & 2.65 & - & 1 & 0.81 \\
    & Kingston Full  & 29.24 & 3   & 0.82 & 0.16 \\
    & Kingston HADOF & 173.77 & 60   & 1 & 0.52 \\
    \hline

    \multirow{3}{*}{30} 
    & Ideal HADOF    & 1.83 & - & 0.97 & 0.78 \\
    & Kingston Full  & 33.74 & 3   & 0.76 & 0.19\\
    & Kingston HADOF & 263.3 & 90   & 0.97 & 0.53 \\
    \hline

    \multirow{3}{*}{40} 
    & Ideal HADOF    & 2.66 & - & 0.95 & 0.86 \\
    & Kingston Full  & 43.4 & 3   & 0.55 & 0.09 \\
    & Kingston HADOF & 306.56 & 120   & 0.94 & 0.49 \\
    \hline

    \multirow{3}{*}{50} 
    & Ideal HADOF    & 4.37 & - & 0.96 & 0.79 \\
    & Kingston Full  & 51.24 & 3   & 0.49 & 0.02 \\
    & Kingston HADOF & 417.32 & 150   & 0.92 & 0.53 \\
    \hline

  \end{tabular}
  \vspace{0.25cm}
\end{table*}

In this comparison, full-circuit QAOA refers to executing the entire optimisation problem as a single parameterised quantum circuit, where all qubits and couplings are encoded into one global ansatz. In contrast, HADOF decomposes the original problem into smaller subproblems, executes these as independent subcircuits, and then combines the results through a federated manner. This structural difference has direct implications for both scalability and performance under noise. A critical implication of this structural distinction is the effective qubit requirement. In a standard full-circuit QAOA formulation, a problem with $n$ binary variables requires $n$ physical qubits, as the entire QUBO must be embedded into a single global circuit. In contrast, HADOF operates on fixed-size subcircuits, allowing the same problem to be solved using only a small number of qubits per execution. In our experiments above, we demonstrated this explicitly by solving QUBO instances with up to 500 variables using only 5 qubit circuits within the HADOF framework. On real IBM hardware, we scaled up to 300 variable problems using the same 5 qubit circuits. This is particularly significant given that the underlying device consists of 156 qubits with limited connectivity; due to hardware topology constraints, embedding dense QUBOs in a monolithic fashion is practically restricted to problem sizes of approximately 50 variables. This highlights a fundamental scalability bottleneck in standard practice full-circuit QAOA: even when sufficient qubits are available in principle, limited qubit connectivity and routing overhead make large dense problem embeddings infeasible. HADOF circumvents this limitation entirely by avoiding global embeddings, enabling both simulation and hardware execution at problem sizes far beyond what is achievable with standard QAOA.

A first important observation from Table \ref{tab:table} is that HADOF consistently delivers stronger solution quality than full-circuit execution, particularly on real hardware. Across all problem sizes, full QAOA exhibits a clear degradation in average accuracy as system size increases, dropping sharply under hardware noise (e.g. down to 0.02 at size 50 on Kingston). In contrast, HADOF maintains significantly higher and more stable average accuracy across all sizes, reaching 0.53 at size 50 on Kingston hardware. This indicates that decomposition effectively mitigates noise accumulation and circuit depth limitations that heavily impact monolithic QAOA circuits.
More strikingly, HADOF on real hardware is often competitive with, and in some cases comparable to, ideal noiseless simulations at this scale. For several problem sizes, the best accuracy (relative score of the best sampled solution) achieved by HADOF on noisy devices approaches or matches the classical simulated annealing baseline. This suggests that the federated structure both stabilises performance under noise and recovers high quality solutions that are difficult to obtain reliably in full-circuit execution even under idealised conditions.

An additional practical constraint is that ideal simulation of full QAOA circuits becomes infeasible beyond approximately 30 qubits on the Mac M3 system due to exponential statevector growth. This limitation restricts meaningful comparison between ideal and hardware runs at larger problem sizes. In contrast, HADOF remains computationally tractable because it operates on smaller subcircuits, allowing both simulation and hardware execution to scale beyond this threshold. As a result, HADOF enables continued benchmarking at problem sizes (40–50 qubits) where full-circuit ideal simulation is no longer possible in practice. In terms of QPU usage, HADOF introduces a predictable overhead relative to full-circuit execution. On ideal simulations, this overhead is moderate and largely linear, reflecting the cost of decomposing and recombining subproblems. However, on real hardware, the overhead becomes more pronounced due to repeated circuit execution, queueing delays, and job submission latency. Despite this, the QPU usage increase does not undermine the quality gains observed, particularly in noisy regimes where full QAOA performance deteriorates significantly. The results indicate that HADOF provides a favourable trade-off between computational overhead and solution quality. While it incurs additional QPU usage costs compared to monolithic QAOA, it achieves substantially improved robustness under noise, maintains strong best case performance across all problem sizes, and extends the range of tractable problem sizes beyond the limits of classical simulation and quantum device constraints.

\subsection{Application to Genome Assembly}

For the application of HADOF to genome assembly, we use the $\phi$X 174 bacteriophage dataset from Boev et al \cite{quantum_genome_assembly} \cite{174bacteriophage}. The genome consists of 5386 base pairs. 50 simulated sequencing reads without errors were created using Grider \cite{grinder} consisting of 600 base pairs each. In their work, the DAG assembly graph was generated and was assembled using Quantum Annealing on D-Wave \cite{dwave-ocean-sdk}. However, the graph was manually partitioned beforehand, so that it can fit on the device due to qubit and connectivity constraints. 

\begin{figure}
    \centering
    \includegraphics[width=0.8\linewidth]{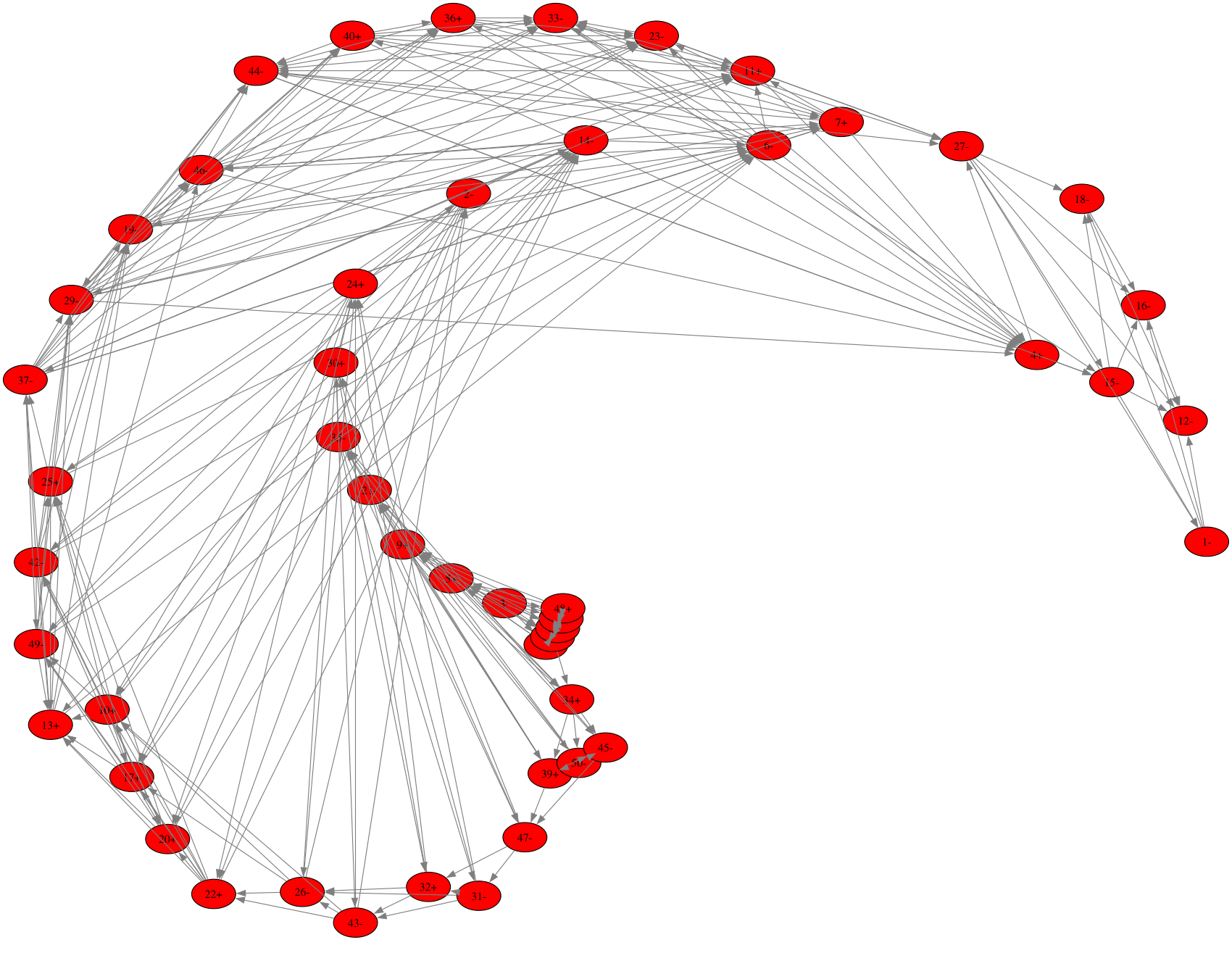}
    \caption{Overlap Graph of $\phi$X 174 from Boev et al \cite{quantum_genome_assembly}}
    \label{fig:overlap}
\end{figure}

\begin{figure}
    \centering
    \includegraphics[width=0.8\linewidth]{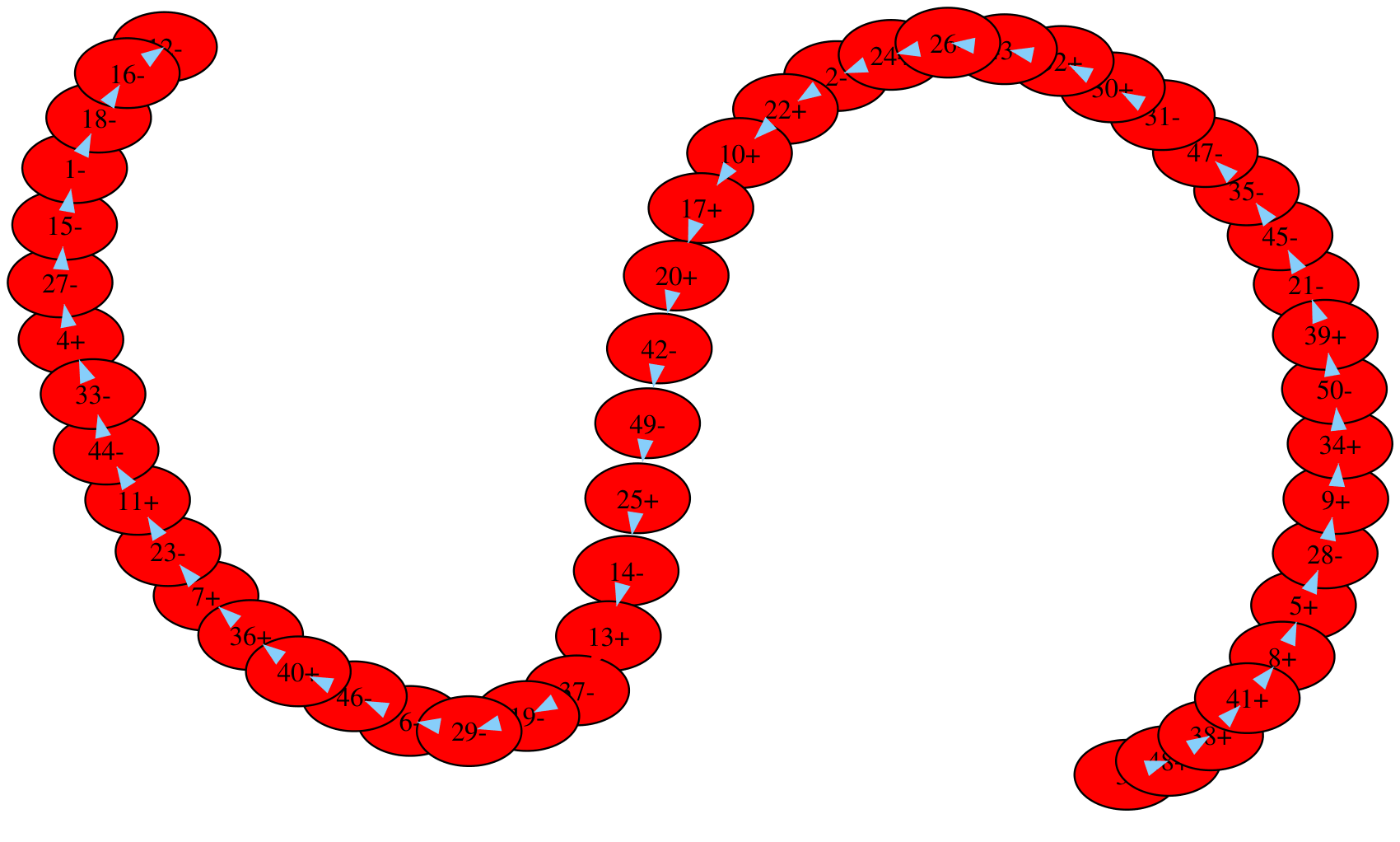}
    \caption{Assembled overlap graph of $\phi$X 174 showing the longest path connecting all the sequencing reads such that they are overlapping and form one contiguous sequence.}
    \label{fig:phix}
\end{figure}

Fig. \ref{fig:overlap} shows the DAG as generated from the dataset. The graph consists of 50 nodes, representing sequencing reads, and 248 edges between them, representing overlaps. Using the QUBO formulation requires 248 binary variables. Using HADOF, we can perform the assembly of this graph in the QUBO form, without requiring any manual partitioning. We compare the results against Simulated Annealing, used as a benchmark. The genome assembly problem is encoded as a QUBO where the optimal solution corresponds to the minimum energy configuration. For this instance, with 248 edges, the ground state energy is $-496$, and since this particular example problem has a unique valid assembly, any deviation from this value directly reflects suboptimal or infeasible reconstructions. Fig. \ref{fig:phix} shows the correct assembly.

\begin{table*}[t]
\vspace{-2em}
  \centering
  \caption{Genome Assembly Solutions}
  \label{tab:assembly}
  \begin{tabularx}{\textwidth}{|l|X|X|X|X|X|}
    \hline
    \textbf{Algorithm} & \textbf{Wall Clock Time} & \textbf{QPU Usage Time} & \textbf{Best Accuracy} & \textbf{Number of Optimal Solutions} & \textbf{Average Accuracy} \\
    \hline
    Simulated Annealing & 2.3 & - & 1 & 213 & 0.99 \\
    Ideal Sequential HADOF & 20.78 & - & 1 & 187 & 0.96 \\
    Ideal Parallel HADOF & 7.23 & - & 1 & 193 & 0.97 \\
    Kingston sequential HADOF & 2344.76 & 750 & 0.91 & 0 & 0.45\\
    Kingston Parallel HADOF & 956.48 & 750 & 0.88 & 0 & 0.51\\
    Multi-QPU Parallel HADOF & 889.04 & 750 & 0.91 & 0 & 0.48\\
    \hline
  \end{tabularx}
\end{table*}

The results in Table~\ref{tab:assembly} provide a clear comparison of how the QUBO based formulation for genome assembly behaves across classical optimisation, ideal quantum simulations, and executions on real quantum hardware. The accuracies are scaled (normalised by $-496$) to allow a consistent comparison across all methods. Classically, Simulated Annealing performs exceptionally well, achieving perfect best accuracy (1.0), nearly perfect average accuracy (0.99), and a high frequency of optimal solutions (213). This indicates that the QUBO landscape for this instance is relatively well behaved, with a sufficiently large basin around the global minimum.

In the ideal, noise-free setting, both sequential and parallel HADOF also recover the optimal solution consistently (best accuracy of 1.0), with high frequencies (187 and 193 respectively). However, a notable distinction emerges in QPU usage: parallel HADOF reduces wall clock time from 20.78 to 7.23 seconds without degrading solution quality. This demonstrates that the QUBO decomposition inherent in HADOF can be effectively parallelised, preserving the integrity of the global energy landscape while improving computational efficiency. The slightly lower average accuracies (0.96–0.97) compared to Simulated Annealing suggest that while the ground state is reliably identified, the sampled distribution is somewhat broader, reflecting differences in exploration dynamics between annealing and variational circuit-based approaches.

The behaviour changes significantly on real hardware (Kingston QPU and multi-QPU configurations), where noise becomes the dominant factor. Although the best accuracies (0.88–0.91) indicate that reasonably good solutions are still reachable, the number of optimal solutions drops to zero across all hardware runs. Despite the QUBO formulation remaining unchanged, the inherent noise in the hardware prevents the system from reliably sampling the true ground state. The degradation in average accuracy (0.45–0.51) further confirms that the distribution of sampled solutions is substantially flattened, with probability mass shifted away from the optimum. Interestingly, parallelisation on hardware (both single-QPU and multi-QPU) still provides clear wall clock advantages (2344.76 down to 956.48 and 889.04), but unlike the ideal case, this does not translate into improved solution quality. The multi-QPU setup achieves similar best accuracy (0.91) to the sequential case, but without any increase in optimal solution frequency, reinforcing that noise remains a limiting factor. Although, optimal solutions were not observed in this example using the real device, it is notable that over 52\% of the solutions in the distribution recovered the correct final sequence by making use GFAtools \cite{gfatools}, a post-processing tools in the genomics pipeline. Further research is required to integrate QUBO based assembly into a standard genome assembly pipeline, making use of the standard pre- and post-processing tools  such as Minimap \cite{Li2018}, GFAtools \cite{gfatools} and Racon \cite{Racon}. 

These results highlight a key distinction in how QUBO based genome assembly behaves across computational paradigms. In classical and idealised quantum simulations, the formulation is highly effective, with all methods consistently identifying the unique optimal solution. However, on current quantum hardware, noise significantly alters the optimisation landscape, reducing both the probability of sampling the ground state and the overall solution quality. However, this research pushes boundaries of the scale of genome assembly problems that can be implemented and optimised, with far higher accuracy than currently available methods.

\subsection{Summary}

This work presents several key advances at the intersection of quantum optimisation and practical quantum computing. This is the first implementation of HADOF on real quantum hardware, moving beyond purely simulated or theoretical studies and demonstrating its behaviour under realistic noise, connectivity, and execution constraints. This is also the first application of HADOF to a real-world problem - genome assembly, where the QUBO formulation captures a structured combinatorial optimisation task with a ground truth solution. This establishes HADOF as a methodological contribution as well as a practical framework for scientific applications. A central contribution of this work is its alignment with the paradigm of HPQ (High Performance Quantum Computing). By design, HADOF enables parallel and federated execution, which translates directly into performance gains across all settings. 

In ideal simulations, parallelisation reduces wall clock time while preserving solution quality. Under noisy simulations and on real hardware, these gains become more pronounced due to the mitigation of execution bottlenecks such as circuit latency and queueing delays. The demonstrated scalability across single and multi-QPU configurations reinforces using HADOF on distributed quantum architectures. Importantly, HADOF achieves superior robustness and accuracy compared to standard full-circuit QAOA, even in regimes where sufficient qubits are available to implement the monolithic approach. While full QAOA suffers from noise accumulation and performance degradation, HADOF maintains consistently higher solution quality by decomposing the problem into smaller, more resilient subcircuits. This highlights that the advantage of HADOF is not merely in enabling larger problem sizes, but also in improving performance within existing hardware limits. This research demonstrates that decomposition and parallelisation are not just scalability tools, but fundamental enablers of practical quantum advantage, positioning HADOF as a strong candidate framework for future large scale quantum optimisation under a HPQ model. 

\section{Conclusions and Future Work}
\label{sec:conclusion}

In this work, we demonstrated HADOF as a scalable framework for solving QUBO problems even on current noisy hardware, including its first real device implementation and its first application to genome assembly. The results show that decomposition and parallelisation enable strong solution quality and robustness to noise, while aligning naturally with the HPQ paradigm for improved makespan time across ideal, noisy, and real hardware settings.

Looking forward, several directions can further strengthen the HADOF framework. From an algorithmic perspective, adaptive and problem specific decomposition strategies are a key avenue for improvement, as prior work has shown that hybrid and decomposition approaches are essential for overcoming hardware limitations in current quantum devices \cite{bass2021decomposition,ponce2023graph}. Recent advances in decomposed and hierarchical QAOA variants further demonstrate that breaking large QUBO problems into smaller subproblems can significantly improve scalability and solution quality \cite{gou2025backbone,minato2024twostep}. In addition, improving circuit design through efficient ans\"atze and optimised circuit depth will be critical, as circuit depth and noise accumulation remain major bottlenecks for QAOA performance \cite{herrman2021depth}. Integrating error mitigation and quantum error detection techniques is another important direction, as these have been shown to substantially improve optimisation performance on noisy quantum hardware \cite{he2025errordetection}. At the systems level, more efficient scheduling, load balancing, and communication across multi-QPU architectures will be necessary to fully realise the HPQ paradigm for distributed quantum optimisation.

On the application side, extending this framework to larger and more complex genome assembly problems presents a natural next step. This includes scaling to higher coverage datasets, incorporating sequencing errors, and refining QUBO encodings to better capture biological constraints such as repeats and ambiguous overlaps. The current dataset was created from simulated sequencing reads. The next step would be to test HADOF on datasets from real DNA sequencers, and integrated HADOF into a standard genome assembly pipeline. HADOF demonstrates that decomposition and parallelisation are not merely workarounds for current hardware limitations, but fundamental design principles for scalable quantum optimisation. By enabling high quality solutions with limited qubit resources and leveraging distributed execution, HADOF provides a practical pathway toward HPQ, where performance gains arise from both algorithmic structure and parallelism.
\section*{Acknowledgement}

We acknowledge the use of IBM Quantum Credits for this work. The views expressed are those of the authors, and do not reflect the official
policy or position of IBM or the IBM Quantum team. All code used in this study will be made publicly available via a GitHub repository upon acceptance of the paper.

\section*{Declaration on Generative AI}

During the preparation of this work, the authors used ChatGPT in order to: Grammar and spelling check, paraphrase and reword. After using this tool, the authors reviewed and edited the content as needed and takes full responsibility for the publication's content. 

\bibliographystyle{unsrt}  
\bibliography{refs}

\end{document}